\documentclass{emulateapj}

\usepackage{epsfig}

\newcommand{\etal}{\mbox{et al.}}
\newcommand{\ergcms}{erg cm$^{-2}$ s$^{-1}$}
\newcommand{\ergs}{erg s$^{-1}$}
\newcommand{\phcms}{ph cm$^{-2}$ s$^{-1}$}

\newcommand{\msun}{$M_{\odot}$}

\newcommand{\chandra}{{\it Chandra}}

\newcommand{\xmm}{{\it XMM-Newton}}

\shortauthors{Muno \etal}
\shorttitle{Diffuse X-rays from Westerlund 1}

\begin{document}

\title{Diffuse, Non-Thermal X-ray Emission from the Galactic Star Cluster Westerlund 1}
\author{
Michael P. Muno,\altaffilmark{1,2}
	Casey Law,\altaffilmark{6} 
        J. Simon Clark,\altaffilmark{3}
	Sean M. Dougherty,\altaffilmark{5}
	Richard de Grijs,\altaffilmark{4}
	Simon Portegies Zwart,\altaffilmark{7,8} and
	Farhad Yusef-Zadeh\altaffilmark{6}
}

\altaffiltext{1}{Department of Physics and Astronomy, University of California,
Los Angeles, CA 90095; mmuno@astro.ucla.edu}
\altaffiltext{2}{Hubble Fellow}
\altaffiltext{3}{Department of Physics \& Astronomy, The Open University, 
Walton Hall, Milton Keynes, MK7 6AA, UK}
\altaffiltext{4}{Department of Physics \& Astronomy, The University of 
  Sheffield, Hicks Building, Hounsfield Road, Sheffield S3 7RH, U.K.}
\altaffiltext{5}{National Research Council, Herzberg Institute of Astrophysics,
   Dominion Radio Astrophysical Observatory, PO Box 248, Penticton BC. V2A 6K3}
\altaffiltext{6}{Department of Physics and Astronomy, Northwestern University, 
   Evanston, IL 60208, USA}
\altaffiltext{7}{Astronomical Institute 'Anton Pannekoek'
        Kruislaan 403, 1098SJ Amsterdam, the Netherlands}
\altaffiltext{8}{Section Computational Science 
        Kruislaan 403, 1098SJ Amsterdam, the Netherlands}

\begin{abstract}
We present the diffuse X-ray emission identified in \chandra\ observations 
of the young, massive Galactic star cluster Westerlund 1. After removing
point-like X-ray sources down to a completeness limit of 
$\approx$$2\times10^{31}$ \ergs, we identify $(3\pm1)\times10^{34}$
\ergs\ (2--8 keV) of diffuse emission. The spatial distribution of the
emission can be described as a slightly-elliptical Lorentzian core
with a half-width half-maximum along the major axis of 
25\arcsec$\pm$1\arcsec,
similar to the distribution of point sources in the cluster, plus a 
5\arcmin\ halo of extended emission. The spectrum of 
the diffuse emission is dominated by a hard continuum component that can 
be described as a $kT$$\ga$3 keV thermal plasma that has a low iron
abundance ($\la$0.3 solar), or as non-thermal emission that could be
stellar light that is inverse-Compton scattered by MeV electrons. 
Only 5\% of the flux is produced by a $kT$$\approx$0.7 keV plasma. The 
low luminosity of the thermal emission and the lack of a 6.7 keV 
iron line suggests that $\la$40,000 unresolved stars with masses between 
0.3 and 2\msun\ are present in the cluster.
Moreover, the flux in the 
diffuse emission is a factor of two lower than would be expected
from a supersonically-expanding cluster wind, and there is no evidence 
for thermal remnants produced by supernovae. 
Less than $10^{-5}$ of the mechanical luminosity of 
the cluster is dissipated as 2--8 keV X-rays, leaving a large amount of 
energy that either is radiated at other wavelengths, is dissipated beyond 
the bounds of our image, or escapes into the intergalactic medium.
\end{abstract}

\section{Introduction}

Sensitive X-ray observations are an increasingly-important tool for 
studying young star clusters, particularly now that the {\it
Chandra X-ray Observatory} and the {\it Newton} X-ray Multi-Mirror Mission
have made harder X-rays (2--10 keV) available for study. Young stars of
all types are strong X-ray sources, with low-mass ($M$$<$3\msun) 
pre-main-sequence stars producing X-rays in their active magnetic 
coronae \citep{pf05,fei05}, and massive OB stars ($M$$\ga$8\msun) 
producing 
X-rays through shocks in their stellar winds \citep{cg91,ber97,ski02a}.
Therefore, using observations of local star forming regions (e.g., Orion) as
templates, measurements of the integrated X-ray luminosities of 
more distant clusters can be used to constrain their total stellar 
population, including the numbers of young stars that may be 
unobservable in the optical and infrared because of extinction or 
source confusion \citep{fei05,ns05}.

X-ray observations of clusters of massive stars  
also reveal diffuse X-ray emission that is produced as stellar winds 
encounter each other and the surrounding interstellar medium 
\citep[ISM; e.g.,][]{sh03,tow03,lyz04,tow05,tow06}. 
Learning the fate of the energy carried by these winds, and eventually by 
supernovae, would provide insight into how galaxies evolve. If the 
energy is transferred to the ISM, it might at first trigger future generations 
of star formation, but a sufficiently large input of energy could clear away 
the ISM and halt star formation. Alternatively, if the energy escapes
a galaxy, stellar winds and supernovae would enrich the intergalactic 
medium with metals.
To determine the fate of that energy, it is necessary to obtain X-ray 
observations of clusters that have a range of ages and 
populations, and that are surrounded by ISM with a variety of densities
\citep[e.g.,][]{tow03,tow05}. 

\begin{figure*}[htb]
\centerline{\epsfig{file=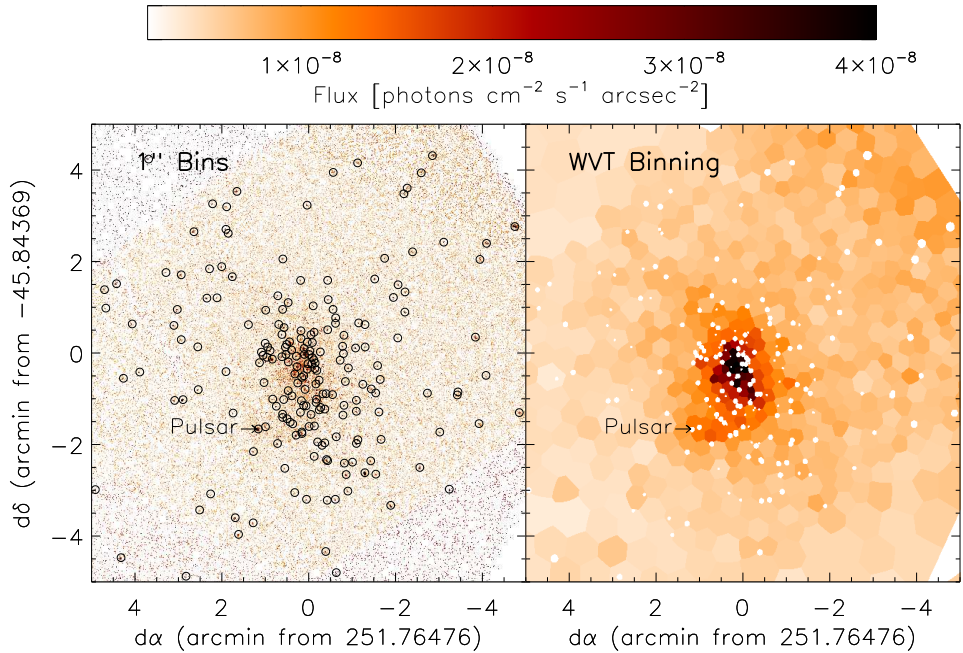,width=0.5\linewidth,angle=90}}
\caption{
Images of the 10\arcmin\ by 10\arcmin\ field around the center
of Westerlund 1. The {\it left panel} displays the unbinned image,
with the point-like X-ray sources described in Clark \etal\ (in prep.)
marked with circles. The {\it right panel} displays an image
of the diffuse X-ray emission in which the point sources have been 
excised and image has been adaptively binned to a signal-to-noise of 
10 using the Weighted Voronoi Tessellation algorithm \citep{ds06}. 
The location of the 10.6 s X-ray pulsar \citep{mun06} is also indicated.
}
\label{fig:diffimg}
\end{figure*}

In this paper, we report on \chandra\ observations of the diffuse X-ray
emission from the young Galactic star cluster Westerlund 1. 
The cluster contains 24 Wolf-Rayet (WR) stars, more than 
80 blue super-giants, at least 3 red super-giants, a luminous blue variable, 
and an amazing 6 yellow hyper-giants, only 6 of which are known in the 
entire rest of the Galaxy \citep{wes87, cn02, cn04, nc05, cla05}.
Assuming a standard initial mass function \citep{krou02}, 
Westerlund 1 could be as massive as $10^5$ \msun, making it
several times larger than the well-known, young Galactic clusters the 
Arches, Quintuplet, and NGC 3603. Westerlund 1 is also 
located only $\approx$5 kpc away 
\citep{cn02,cla05}, so it is one of the closest young, dense star 
clusters. Therefore, Westerlund 1 is 
a crucial object for understanding the evolution of star clusters, and their
impact on the ISM of their host galaxies.

This is one of several papers describing the \chandra\ observations.
In \citet{mun06}, we reported the detection of a slow X-ray pulsar in 
Westerlund 1. \citet{ski06} examined the X-ray emission from the 
Wolf-Rayet (WR) stars in the cluster, as well as a subset of the OB supergiants
that are brightest in X-rays. 
In Clark \etal\ (in prep.), we will report the spectroscopic identification of 
further optical counterparts to the X-ray sources, and discuss the origin
of the X-ray emission from these stars. 
In this paper, we describe the spatial distribution (\S\ref{sec:dist})
and spectrum (\S\ref{sec:spec}) of the diffuse X-rays. We compare the
emission seen from Westerlund 1 to that of other massive star clusters in
the Local Group (\S3). We suggest that Westerlund 1 is one of only a few
star clusters to produce mostly non-thermal X-rays. We discuss the contraints
that this places on the contributions of pre-main-sequence stars (\S3.1), 
stellar winds (\S3.2), and supernovae (\S3.3) to the diffuse emission, 
and examine what could be causing non-thermal emission (\S3.4). 

\section{Observations and Data Analysis\label{sec:obs}}

Westerlund 1 was observed with the {\it Chandra X-ray Observatory}
Advanced CCD Spectrometer Spectroscopic array \citep[ACIS-S;][]{wei02} 
on two occasions: on 2005 May 22 for 18 ks (sequence 5411), and on 
2005 June 20 for 42 ks (sequence 6283). We reduced the observation using
standard tools that are part of CIAO version 3.3. We first created a 
composite event list for each observation. We corrected the pulse heights 
of the events to mitigate for the position-dependent charge-transfer 
inefficiency using the standard CIAO process and excluded 
events that did not pass the standard ASCA grade filters and \chandra\ X-ray
center (CXC) good-time 
filters. We searched for intervals during which the background 
rate flared to $\ge 3\sigma$ above the mean level, and removed one such
interval lasting 3.6 ks from sequence 5411. The composite image of the full
field is displayed in Clark \etal\ (in prep.); Figure~\ref{fig:diffimg}
displays the inner 10\arcmin\ of the cluster at 1\arcsec\ resolution.
 
As described in Clark \etal\ (in prep.), we identified point-like X-ray sources 
in each observation using a wavelet-based algorithm, wavdetect \citep{free02}.
%
In order to examine the diffuse X-ray emission, we then removed events that 
fell within circles circumscribing approximately 92\% of the PSF at the 
location of each point source, and created an image using the remaining 
photons. Within 5\arcmin\ of the cluster core, 7386 counts were 
associated with known point sources, and 38,350 counts in diffuse emission,
so photons from point sources in the wings of the point spread function
contribute only 0.5\% to the diffuse flux. 


\begin{figure}
\centerline{\epsfig{file=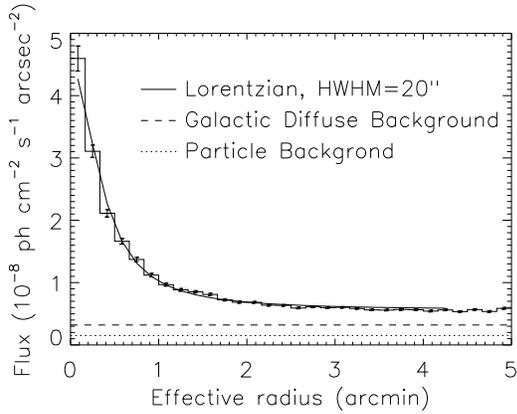,width=0.95\linewidth}}
\caption{
The radial profile of the diffuse X-ray emission (histogram with
errors) along with the best-fit model described by Equation~\ref{eq:prof}.
}
\label{fig:diffdist}
\end{figure}

\begin{deluxetable}{lc}
\tablecolumns{2}
\tablewidth{0pc}
\tablecaption{Lorentzian Model of the Distribution of Diffuse X-rays\label{tab:diffdist}}
\tablehead{
\colhead{Parameter} & \colhead{Value} \\
} 
\startdata
$C$ $10^{-8}$ \phcms & 0.53$\pm$0.04 \\
$N$ $10^{-8}$ \phcms & 3.4$\pm$0.2 \\
$\alpha_0$ (J2000) & 16 47 4.3$\pm$0.1 \\
$\delta_0$ (J2000) & $-$45 50 59$\pm$1 \\
$r_0$ (arcsec) & 25$\pm$1 \\
$\epsilon$ & 0.75$\pm$0.02 \\
$\theta$ (degrees) & 13$\pm$3 \\
$\chi^2/\nu$ & 360/120 
\enddata
\tablecomments{We define $\theta$ as positive for rotations east of north.}
\end{deluxetable}

\subsection{Spatial Distribution\label{sec:dist}}

The signal-to-noise in a 1\arcsec\ pixel
was low, so we adaptively binned the image
using the Weighted Voronoi Tessellation algorithm implemented by 
\citet{ds06}, which is based on the algorithm of 
\citet[][]{cc03}. The resulting image is displayed in
the right panel of Figure~\ref{fig:diffimg}.

In order to quantify the extent of the diffuse emission, we modeled its 
adaptively-binned, two-dimensional spatial distribution 
(Figure 1) as a Lorentzian function. 
Other functional forms used to model
the light from star clusters also may be consistent with the data 
\citep[e.g.,][]{eff87,agg06}. However, there is little tradition in modeling
the spatial distribution of diffuse X-rays from star clusters with 
analytic functions, because that emission usually has a complex
morphology \citep[e.g.,][]{tow03}, so there is not an obvious choice
for a functional form. We chose a Lorentzian function 
for its simplicity, and because it is similar to the 
King models often used to quantify the distribution of optical light from 
star clusters \citep{kin62}.
The diffuse emission from Westerlund 1 is not circularly symmetric, so we 
allowed for an elliptical distribution defined as 
\begin{eqnarray}
f(x^\prime, y^\prime) = C + {{N}\over{1 + ({x^\prime}^2 + \epsilon^2 {y^\prime}^2 )/r_0^2}}, \\
\nonumber x^\prime = (\alpha - \alpha_0)\cos\delta_0 \cos\theta + (\delta - \delta_0)\sin\theta \\
\nonumber y^\prime = (\delta - \delta_0)\cos\theta - (\alpha - \alpha_0)\cos\delta_0 \sin\theta
\label{eq:prof}
\end{eqnarray}
Here, $\alpha_0$ and $\delta_0$ are the center of the distribution
and $x^\prime$ and $y^\prime$ are the offset in arcseconds from the 
center, where the axes defining them have been rotated 
east of north by $\theta$ degrees. The remaining parameters are the 
background count rate $C$, the peak count rate $N$, the ellipticity of the
distribution $\epsilon$ (a value of 1 implies a circle), and the characteristic
radius of the distribution $r_0$. The final parameters and the 
goodness-of-fit $\chi^2/\nu$ are listed in Table~\ref{tab:spec}. 

The fit is formally poor, with $\chi^2/\nu$$=$3, because there are 
structures in the diffuse emission that can not be described as 
part of an elliptical Lorentzian, including some bright knots of emission
at the center of the cluster, and a ridge of emission extending to
the southeast toward the X-ray pulsar reported by 
\citet[][it is also labeled in Fig.~\ref{fig:diffimg}]{mun06}. 
However, the model does provide a useful means to quantify the 
azimuthally-averaged distribution of the emission, as can be seen 
in the radial distribution, plotted in units
of $r^\prime = ({x^\prime}^2 + \epsilon^2{y^\prime}^2)^{1/2}$,
in Figure~\ref{fig:diffdist}. 
The half-width half-maximum of the
distribution is 25$\pm$1\arcsec, which for a distance of 5 kpc corresponds
to 0.5 pc. The widths of the distributions of both optical
stars and point-like X-ray sources are also $\approx$25\arcsec\ 
\citep{cla05,mun06} . Moreover, the centroid of the diffuse 
emission lies within the 5\arcsec\ uncertainty in the centroid of
the point sources, \citep{mun06}, although it is $\approx$20\arcsec\ 
from the cetroid of the optically-detected stars 
\citep[][]{pbc98, cla05}. The discrepancy 
could be caused either by differential extinction toward the cluster, or
by substructure in the cluster (J. S. Clark \etal, in prep.).  

In Figure~\ref{fig:diffdist}, we have also indicated the 
amount of flux expected from the background of particles impacting
the detector ($1.5\times10^{-9}$ \phcms), and the mean flux taken from 
observations of the Galactic plane at $l = 28^\circ$ and $b = 0.2^{\circ}$ 
\citep[$2.2\times10^{-9}$ \phcms, of which half was 
particle background; see][]{ebi05}. Even 5\arcmin\ from the cluster
core, 60\% of the flux was from Westerlund 1, so there is a broad halo of 
diffuse X-rays around the cluster. In contrast, although a few O stars
that are cluster members are located beyond $\sim$3\arcmin\ from the 
cluster core \citep{cla05}, the surface density of X-ray point sources 
beyond 3\arcmin\ is consistent that of the Galactic disk \citep{mun06}, 
so there is no similar halo of point-like X-ray sources.

\subsection{Spectra\label{sec:spec}}

\begin{figure*}[htb]
\centerline{\epsfig{file=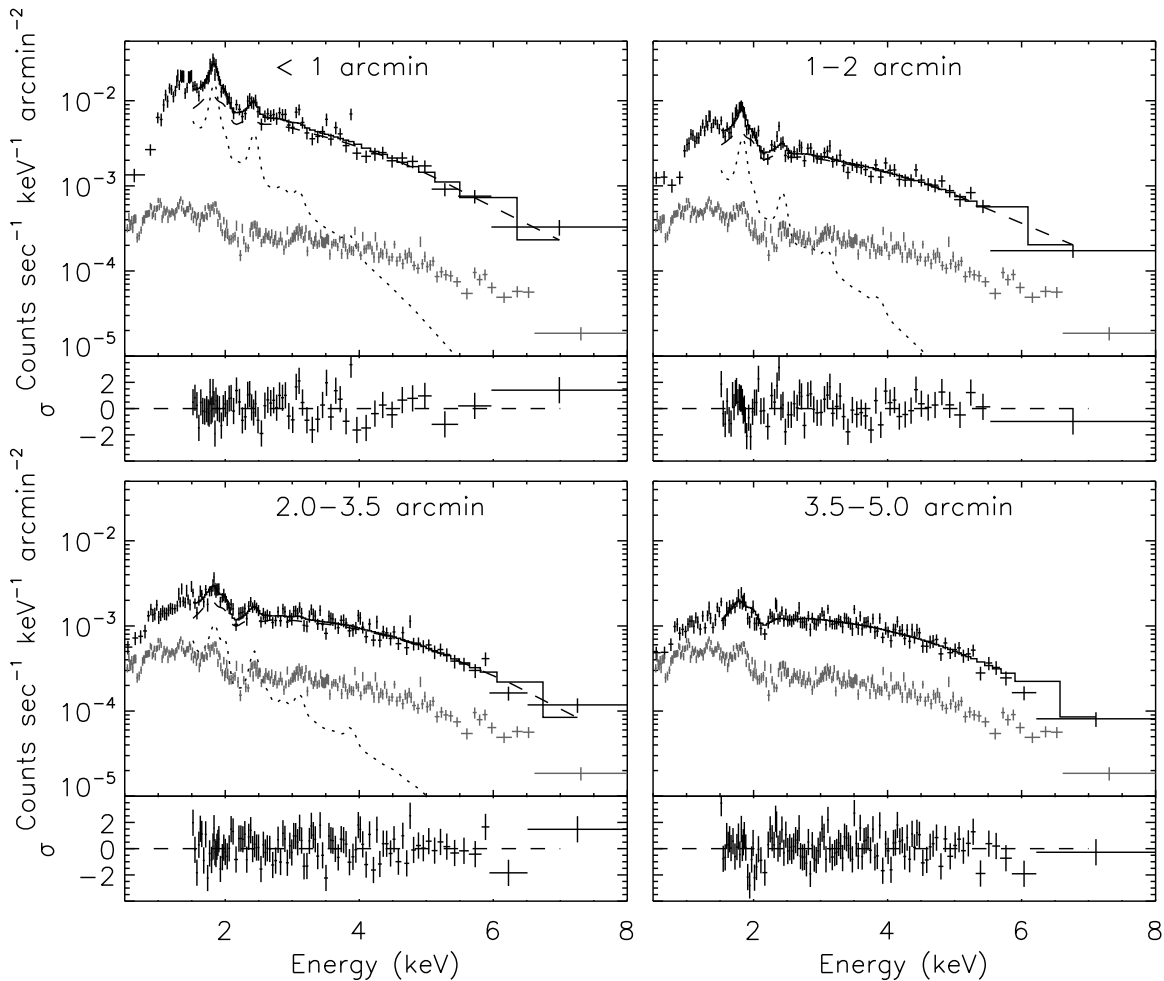,width=0.95\linewidth}}
\caption{
Background subtracted spectra of the diffuse X-ray emission from Westerlund 1.
In the top half of each panel, the spectra are plotted in detector units, 
so the intrinsic spectrum is still convolved with the response of the detector
and telescope. The model 
spectrum is plotted with the solid histogram. The dotted line denotes
the thermal component of the model, and the dashed line the non-thermal
power-law component. The spectrum becomes noticeably harder farther from
the cluster, because the thermal component becomes weaker. The grey data
points represent Galactic plane emission at $l = 28^\circ$ and 
$b = 0.2^{\circ}$, which contributes only 30\% of the total flux toward
Westerlund 1. In the
bottom half of each panel, we plot the difference between the data and model
normalized to the uncertainty ($\sigma$) in the data.
}
\label{fig:diffspec}
\end{figure*}

Guided by Figure~\ref{fig:diffdist}, we extracted spectra, 
response functions, and effective area curves
for a circular 1\arcmin\ region around the cluster center, and annular 
regions 1--2\arcmin, 2--3.5\arcmin, and 3.5--5\arcmin\ from the cluster 
center. The
background-subtracted spectra are displayed in Figure~\ref{fig:diffspec},
in units of detector counts per square arcminute.

The spectra contained
contributions from at least three sources of X-rays: plasma and unresolved
stars in the cluster, diffuse emission from the Galactic plane, 
and events produced by particles incident on the detectors. 
The spectrum 
of the background from particles has been well-characterized using 
observations in which the ACIS detectors were stowed out of the focal plane.
Therefore, we subtracted the spectrum of the particle background from 
our source spectra. However, we were not able to make a local
estimate of the Galactic emission, because the cluster
emission extended over the entire image. We have
not attempted to subtract the Galactic plane emission from the spectra 
of the diffuse emission from Westerlund 1, but have estimated
the contribution of the Galactic emission by modeling a spectrum from 
observations of a field at $l = 28^\circ$ and $b = 0.2^{\circ}$ 
\citep{ebi01,ebi05}. 

We modeled the emission using
version 12.2.0 of XSPEC \citep{arn96}. We chose to model only the
1.5--8 keV energy range, for two reasons. First, the mean absorption 
column measured from the X-ray spectra of the 
point sources in Westerlund 1 is equivalent to $2\times10^{22}$ cm$^{-2}$
of hydrogen (J. S. Clark \etal, in prep.). Therefore, most of the 
observed X-ray
flux from the cluster should emerge at energies above $\sim$2 keV, and 
the lower-energy X-rays are probably from foreground emission. 
Second, given that the diffuse emission
from Westerlund 1 is probably from several different sources, we do not have 
enough physical guidance to extrapolate our models below 2 keV. 
When we try to apply simple models, the inferred
de-reddened 0.5--2.0 keV flux can span an order-of-magnitude 
depending upon the model assumptions. The lower bound was  
chosen to include the prominent line at 1.8 keV in the spectra from the 
inner 2\arcmin\ of the cluster. For 
consistency with the other works quoted in \S3,
we report the observed and de-reddened 2--8 keV fluxes.

We first attempted to model the spectrum as a single temperature 
thermal plasma, either in or out of collisional ionization equilibrium,
absorbed by the interstellar medium. We found that those simple models 
provided a poor description of the data from the inner 2\arcmin\ of 
the cluster ($\chi^2/\nu \ge 1.5$).

\begin{deluxetable*}{lcccc}[htb]
\tablecolumns{2}
\tablewidth{0pc}
\tablecaption{Two-Temperature Plasma Model for the Diffuse X-ray Emission\label{tab:spec}}
\tablehead{
\colhead{Parameter} & \colhead{$<$1\arcmin}  & \colhead{1--2\arcmin}  & \colhead{2.0-3.5\arcmin} & \colhead{3.5-5.0\arcmin} \\
} 
\startdata
counts & 4259 & 6679 & 12698 & 14714 \\
background & 698 & 2244 & 5613 & 6636 \\
area (arcmin$^{2}$) & 2.7 & 8.9 & 25.6 & 34.9 \\ [5 pt]
$N_{\rm H}$ ($10^{22}$ cm$^{-2}$) & 2.2$^{+0.2}_{-0.2}$ & $2.0^{+0.2}_{-0.2}$ & $2.2^{+0.2}_{-0.2}$ & $2.4^{+0.1}_{-0.1}$ \\
$kT_1$ (keV) & $0.71^{+0.08}_{-0.12}$ & $0.7^{+0.1}_{-0.1}$ & $1.0^{+0.2}_{-0.2}$ & \nodata \\
$Z_1/Z_\odot$ & 2.0 & 2.0 & 2.0 & \nodata \\
$\int n_{e,1} n_{H,1} dV$ (cm$^{-6}$ pc$^{3}$) & $9^{+5}_{-2}$ & $6^{+2}_{-1}$ & $6^{+2}_{-2}$ & \nodata \\ 
$kT_2$ (keV) & $3.2^{+0.5}_{-0.4}$ & $5.7^{+1.2}_{-0.8}$ & $11^{+4}_{-2}$ & $6.3^{+0.9}_{-0.8}$ \\
$Z_2/Z_\odot$ & $<$0.4 & $<$0.3 & $0.6^{+0.5}_{-0.4}$ & $<$0.2 \\
$\int n_{e,2} n_{H,2} dV$ (cm$^{-6}$ pc$^{3}$) & $16^{+3}_{-2}$ & $17^{+1}_{-2}$ & $24^{+2}_{-2}$ & $43^{+1}_{-3}$ \\ 
$\chi^2/\nu$ & 73.4/65 & 80.4/72 & 104.9/101 & 129.1/116 \\ [5pt]
$F_{\rm X}$ ($10^{-13}$ \ergcms) & 7.1$\pm$0.9 & 11.3$\pm$1.0 & 23.8$\pm$2.0 & 27.0$\pm$2.0 \\
$uF_{\rm X,1}$ ($10^{-13}$ \ergcms) & 1.5 & 0.9 & 2.0 & \nodata \\
$uF_{\rm X,2}$ ($10^{-13}$ \ergcms) & 8.3 & 13.4 & 27.7 & 34.9
\enddata
\tablecomments{Uncertainties are 1$\sigma$, found by varying each 
parameter until $\Delta\chi^2 = 1.0$. $uF_{\rm X,1}$ and
$uF_{\rm X,2}$ are the de-absorbed 2--8 keV flux from the thermal
and non-thermal components of the spectral model, respectively. If 
we extrapolate our models to the 0.5--2.0 keV band, the 
rapidly-increasing contribution from the $kT$=0.7 keV
thermal plasma causes the inferred X-ray luminosity to be a factor of
2--3 larger. Note also that at 5 kpc, 1 arcmin = 1.45 pc.}
\end{deluxetable*}

\begin{deluxetable*}{lcccc}[htb]
\tablecolumns{2}
\tablewidth{0pc}
\tablecaption{Thermal Plus Non-Thermal Model for the Diffuse X-ray Emission\label{tab:pow}}
\tablehead{
\colhead{Parameter} & \colhead{$<$1\arcmin}  & \colhead{1--2\arcmin}  & \colhead{2.0-3.5\arcmin} & \colhead{3.5-5.0\arcmin} \\
} 
\startdata
$N_{\rm H}$ ($10^{22}$ cm$^{-2}$) & 2.7$^{+0.3}_{-0.2}$ & 2.3$^{+0.3}_{-0.2}$ & 2.4$^{+0.2}_{-0.2}$ & 2.8$^{+0.3}_{-0.2}$ \\
$kT_1$ (keV) & $0.68^{+0.10}_{-0.08}$ & $0.68^{+0.09}_{-0.13}$ & $1.1^{+0.2}_{-0.1}$ & \nodata \\
$Z_1/Z_\odot$ & 2.0 & 2.0 & 2.0 & \nodata \\
$\int n_e n_H dV$ (cm$^{-6}$ pc$^{3}$) & $11^{+4}_{-3}$ & $7.1^{+0.8}_{-1.8}$ & $6^{+2}_{-2}$ & \nodata \\ 
$\Gamma$ & 2.7$^{+0.2}_{-0.2}$ & 2.1$^{+0.1}_{-0.2}$ & 1.7$^{+0.1}_{-0.1}$ & 2.0$^{+0.2}_{-0.1}$ \\
$N_\Gamma$ ($10^{-3}$ ph cm$^{-2}$ s$^{-1}$ keV$^{-1}$) & $1.0^{+0.3}_{-0.3}$  & $0.7^{+0.2}_{-0.1}$ & $0.8^{+0.1}_{-0.1}$ & $1.7^{+0.4}_{-0.2}$ \\
$\chi^2/\nu$ & 70.4/66 & 80.4/73 & 107.2/102 & 132.2/116 \\ [5pt]
$F_{\rm X}$ ($10^{-13}$ \ergcms) & 7.2$\pm$0.7 & 11.7$\pm$1.0 & 23.1$\pm$2.0 & 28.0$\pm$2.0 \\
$uF_{\rm X,1}$ ($10^{-13}$ \ergcms) & 1.6 & 1.0 & 2.2 & \nodata \\
$uF_{\rm X,2}$ ($10^{-13}$ \ergcms) & 9.0 & 14.2 & 27.2 & 37.3
\enddata
\tablecomments{See Table 2.}
\end{deluxetable*}

Therefore, we modeled the spectra as an absorbed, 
two-temperature thermal plasma. Most of the 4--8 keV continuum flux could 
be modeled as a hot plasma with $kT_2$$\ga$ 3 keV.
In the inner 3.5\arcmin\ of the cluster, the presence of emission 
lines near 1.8 keV from He-like Si and 2.3 keV from He-like S indicated 
that some of the flux is produced by a cool $kT_1$$\la$1 keV plasma.
The metal abundances in the 
cooler component were poorly constrained because it contributes very
little to the continuum emission, so we fixed the metal abundances to 
the mean best-fit value of $Z/Z_\odot$=2. 
Moreover, the spectrum taken from the 3.5--5.0\arcmin\ annulus 
lacked obvious line emission, so the parameters of any cool plasma 
emission were unconstrained. Therefore, we omitted the
cool component from the model of that spectrum.
In Table~\ref{tab:spec}, we list the parameters of the best-fit, two
temperature, collisional ionization equilibrium models; using non-equilibrium
models yields similar results for the derived temperatures and abundances. 
Using these assumptions, the models were 
generally good descriptions of the data, with $\chi^2/\nu$$\approx$1.

The most notable trend in the cool components is that their contributions
to the spectra decline from 15\% in the central 1\arcmin, to 7\% between
1\arcmin\ and 3.5\arcmin, finally becoming undetectable beyond 3.5\arcmin\
from the cluster center. Otherwise, the inferred interstellar absorption
remains roughly constant near $2.6\times10^{22}$ cm$^{-2}$, the temperature
of the thermal component is constant near $kT$$\approx$$0.7$--1 keV.
We find that the temperature of the hot plasma increases 
from $kT_2$=3 keV at the cluster center, to a maximum 11 keV in the
2.0--3.5\arcmin\ annulus, and then decreases to 6 keV in the 
outer annulus. The relative lack of flux near the He-like Fe line at 6.7 keV 
in most of the spectra implies that the iron abundances are less than 
half of the solar values. Interestingly, similar sub-solar iron abundances
are inferred from the lack of 0.8--1.0 keV Fe L lines from several known 
Galactic Wolf-Rayet and O stars \citep[e.g.,][]{ski01,ski02a,sch03,ski05}. 

Alternatively, the lack of lack of line emission near 6.7 keV could 
be explained if much of the continuum X-ray emission is non-thermal. 
Therefore, we have also modeled the emission as the sum of emission from 
a $kT$$\la$1 keV thermal plasma and a power law (Table~\ref{tab:pow}). 
The metal abundances in the cooler component were once again fixed to 
$Z/Z_\odot$=2, and we omitted the cool component from the model of the
3.5--5\arcmin\ annulus. This provides an equally good description of the
data as the two-temperature plasma model, and the same trends are evident:
the contribution of the cool component declines monotonically with offset
from the cluster center, and the overall spectrum becomes harder. 

The contributions of each model component under the second set of models
are indicated in Figure~\ref{fig:diffspec}, 
using dotted lines for the thermal plasma, and 
dashed lines for the power law component. 
We also display the spectrum of the Galactic
ridge emission at $l = 28^\circ$ and $b = 0.2^{\circ}$ (grey data points). 
The line emission from the Galactic flux is a bit stronger than that
from Westerlund 1, but otherwise the spectra are fairly similar. Therefore,
we cannot completely rule out the hypothesis that the emission beyond 
$\approx$2\arcmin\ from the cluster core is Galactic. However,
our assumption that the diffuse emission is from Westerlund 1 is 
conservative. As described in \S3, we find that the luminosity of
diffuse X-rays from Westerlund 1 is much lower than expected, and  
assuming that the diffuse halo is Galactic would exacerbate the 
discrepancy.

The total, de-reddened
2--8 keV flux from within 5\arcmin\ of Westerlund 1 is 
$9.3\times10^{-12}$ \ergcms. By varying the assumptions in our model, we 
find that the systematic uncertainty in the 2--8 keV flux is
$\approx$20\%. Based on the \chandra\ observations taken at $l = 28^\circ$ and 
$b = 0.2^{\circ}$ \citep[see also][]{ebi05}, we expect the Galactic emission 
to be $3\times10^{-14}$ \ergcms\ arcmin$^{-2}$ 
(2--8 keV; see also Hands \etal\ 2004). Therefore, within 5\arcmin\ of the 
core of Westerlund 1 
the Galactic plane contributes $\approx$20\% to the inferred flux. 
Subtracting this foreground and background emission, and using a distance
to Westerlund 1 of 5 kpc \citep{cla05}, we find that the luminosity of the
diffuse X-ray emission from the cluster is $(3\pm1)\times10^{34}$ \ergs\ 
(2--8 keV). Only $\approx$5\% of this luminosity is from the $\la$1 keV 
thermal component.

\section{Discussion}

The origin of the diffuse X-ray emission from clusters of massive
young stars is currently under debate. Several authors 
\citep[e.g., Cant\'{o}, Raga, \& Rodr\'{i}guez 2000;][]{sh03}
have modeled the diffuse X-rays from the most massive clusters in the
Local Group as a cluster wind. Under this model, the winds of individual 
stars collide, thermalize, and form a pressure-driven bulk 
flow that expands supersonically into the interstellar medium \citep{cc85}.
\citet{sh03} tabulated results from studies of R136, 
NGC 3603 \citep{mof02}, NGC 346 \citep{naz02}, the Rosette \citep{tow03},
and the Arches \citep{yz02}, and showed that the luminosities of their
diffuse X-ray emission ($[1-6]\times10^{34}$ \ergs) were considerably 
larger than would be expected from the standard cluster wind model. 
The large X-ray luminosities can be explained several ways: the densities of 
the cluster winds could be higher than expected because the stellar 
winds entrained cooler material or because radiative losses decreased the
temperature of the shocked plasma \citep{sh03}; the wind energy could
be dissipated through heat conduction where it encounters nearby molecular 
clouds \citep{dm87}; or the wind could be confined by the surrounding ISM 
\citep{chu95}. Alternatively, the large X-ray luminosities might partly 
result from the fact
that unresolved pre-main-sequence stars should contribute significantly to 
the luminosity of the (apparently) diffuse emission, especially for
more distant clusters \citep[e.g.,][]{tow06}. 

Westerlund 1 is at least as massive as NGC 3603, R136, and the 
Arches \citep{cla05}, so from an observational standpoint the luminosity of 
its diffuse X-rays ($[3\pm1]\times10^{34}$ \ergs; 2--8 keV) is understandable. 
However, the spectrum and spatial distribution of the diffuse X-ray emission 
from Westerlund 1 presents more of a puzzle. First, the spectrum lacks the 
line emission from He-like Fe that would be expected from a thermal plasma
given the hard continuum flux. 
In contrast, the spectra of pre-main sequence stars exhibit prominent 
lines from He-like and H-like Si, S, Ar, 
and Fe that imply metal abundances up to ten times the solar value 
\citep[e.g.,][]{fei05}. However, X-ray spectra of O and WR stars often 
exhibit weak Fe emission that imply abundances $\la$0.3 solar 
\citep[e.g.,][]{ski01,ski02a,sch03,ski05}, so it is possible that the diffuse
flux is dominated by plasma from the O and WR star winds, with little 
contribution from pre-main-sequence stars. It is also possible that the
diffuse emission is non-thermal, by analogy with similar interpretations
for the hard flux from a handful of young stellar associations, including 
RCW 38 \citep{wol02}, DEM L192 \citep[N51D;][]{coo04}, 30 Dor C 
\citep{bam04}, and possibly the Arches cluster \citep{lyz04}. 
In most of the above cases, the non-thermal 
emission has been interpreted as synchrotron emission from supernova 
remnants. We will examine this hypothesis for Westerlund 1 in 
\S\ref{sec:nonthermal}.

The second surprise is that the the diffuse X-ray emission from Westerlund 1 
seems to extend far beyond the core of the cluster. Within the inner 
2\arcmin, the surface brightness of the diffuse emission falls off with 
a half-width half-maximum of 0.5 pc (Fig.~\ref{fig:diffimg} and 
\ref{fig:diffdist}), which is identical 
to the distribution of point-like X-ray sources (Clark \etal, in prep). This
core of diffuse emission could be produced either from the cluster wind, which
radiates X-rays mostly in the region where the colliding winds are thermalized
\citep{sh03}, or from an unresolved population of pre-main sequence stars. 
However, between 2\arcmin\ and 5\arcmin\ (3--7 pc) from the cluster 
core the diffuse X-ray flux attains a constant level
$\approx$$7\times10^{-14}$ \ergcms\ arcmin$^{-2}$ (Tab. \ref{tab:spec};
Fig. \ref{fig:diffimg}), which is 2--3 times larger than is expected
from the Galactic plane \citep[e.g.,][]{han04,ebi05}. An expanding thermal
plasma would exhibit a rapidly-declining temperature profile, yet the spectrum
of this halo of diffuse emission is quite hard and lacks the line emission 
expected from a cooling plasma. This makes it tempting to interpret the 
diffuse halo as non-thermal particles that are accelerated in a 
large-scale outflow.

Therefore, although the luminosity of the diffuse X-ray emission from 
Westerlund 1 is not surprising, the lack of line emission in the spectrum 
and broad spatial distribution of the diffuse X-rays is. To address 
this, in the 
following sections we quantify the probable contributions of pre-main-sequence
stars, stellar winds, and supernovae to the X-ray emission from Westerlund 1.

\subsection{Unresolved Low-Mass Stars}

The non-thermal spectrum of the diffuse X-ray emission from Westerlund 1
puts interesting constraints on the population of low-mass stars in the 
cluster. The average spectrum of the lightly-absorbed pre-main-sequence
stars in Orion can be qualitatively described as a two-temperature 
plasma with $kT_1$=0.5 keV and $kT_2$=3.3 keV, and with metal abundances of up
to 10 times solar for S, Ar, Ca, and Fe \citep{fei05}. 
In contrast, in our models for the diffuse emission from Westerlund 1,
a $kT$$\la$1 keV plasma contributes only 5\% of the 2--8 keV diffuse X-ray 
flux, and the remaining hard flux does not exhibit the expected He-like Fe
line at 6.7 keV, placing an upper limit on the Fe abundance of $\la$0.3 solar
(Tab.~\ref{tab:spec}). To obtain a conservative estimate
of the number of low mass stars in the cluster, we assume that they 
have solar Fe abundances (i.e., much lower abundances than the stars in 
Orion). We find that implies that they they produce $\la$30\% of the diffuse 
flux from Westerlund 1, or $\approx$$9\times10^{33}$ \ergs (2--8 keV).

We use the results of the \chandra\ Orion Ultradeep Project (COUP) to 
convert this luminosity into a number of low-mass stars, taking into
account the difference in ages between the two clusters. 
For Orion, the 1398 stars later than B4 in the COUP observations have an 
integrated, de-absorbed 2--8 keV luminosity of $1.2\times10^{33}$ \ergs\
\citep{fei05}. Most of this emission is produced by stars with 
0.3$<$$M$$<$3 \msun.\footnote{The COUP sample of X-ray sources is complete 
to $\approx$0.2\msun, and excluding the singularly bright O6 star 
$\theta^1$ Ori C (with a luminosity of 
$3\times10^{32}$ \ergs\ [2--8 keV], it could be detected as a point source in 
our observations of Westerlund 1) the $>$3\msun\ stars in Orion produce
only $\approx$6\% of the integrated X-ray luminosity.} However, the stars in 
Westerlund 1, with ages of $\approx$4 Myr \citep{cla05}, are significantly 
older than the 1-Myr-old population in Orion. To take this into account,
we first note that when 2--3\msun\ stars reach an age of $\approx$4 Myr, 
they become fully 
radiative and their X-ray luminosities drop by an order of magnitude
\citep{fla03}. Even though they are only 5\% of low-mass stars by 
number, in Orion these 2--3\msun\ stars produce $\approx$30\% of the
 flux from 0.3$<$$M$$<$3 \msun\ stars 
\citep[see Fig 4.\ in][]{fei05}, or $\approx$$8\times10^{32}$ \ergs. 
Second, \citet{pf05} find that the X-ray luminosities
of young stars with 0.5$<$$M$$<$2\msun\ fall off with time $\tau$ as
$L_{\rm X} \propto \tau^{-0.75}$, so at 4~Myr the stars in Orion
should be $\approx$3 times fainter. Therefore,
if Orion were 4 Myr old, we would expect its $\approx$1400 stars with 
0.3$<$$M$$<$2 \msun\ to have a luminosity of $3\times10^{32}$ \ergs\
(2--8 keV). Our upper limit to the integrated X-ray luminosity of 
low-mass stars in Westerlund 1 is $\la$$9\times10^{33}$ \ergs, so
we infer that Westerlund 1 contains $\la$40,000 
stars with masses between 0.3$<$$M$$<$2\msun.

This number of low-mass stars is smaller than one would expect if one
were to extrapolate from the number of massive, post-main-sequence stars
in the cluster using a standard initial mass function \citep{krou02}.
There are $\approx$150 stars brighter 
than $V$=21 within 5\arcmin\ of the center of Westerlund 1, the 
faintest of which have recently been identified as O7 main sequence 
stars that would have initial masses $\ga$30 \msun\ 
(J. S. Clark \etal\ in prep). The maximum initial mass 
of the stars remaining in the cluster is uncertain because there is no 
precise means to determine the initial masses of the supergiants and WR stars,
but \citet{cla05} argue that it is probably in the range of 40--50 \msun.
If we assume that the initial mass function can be described as a broken
power law of the form $dN \propto M_i^{-\alpha} dM_i$, where 
$\alpha$=2.3 for $M_i$$>$0.5\msun\ and $\alpha$=1.4 for 
0.3$<$$M_i$$<$0.5 \msun\ \citep{krou02}, then if there are 150 stars with 
30$<$$M_i$$<$50 \msun, we would expect 100,000 stars with 
0.3$<$$M$$<$2 \msun. This inferred lack of low-mass stars can be 
explained several ways. The slope of the initial mass function could be flat
($\alpha$$\la$2.1), as has been inferred for NGC 3603 and the Arches cluster
based on infrared star counts \citep[e.g.,][]{eis98,sto05}. The mass
function could be truncated at low masses ($M$$<$0.6\msun), by 
analogy with the fact that $M$$<$7\msun\ stars appear to be depleted in 
the Arches cluster \citep{sto05}. Finally, the initial 
masses of the post-main sequence stars in Westerlund 1 could span a 
much wider range of masses (20--60\msun) than assumed in \citet{cla05}. 

If we assume that the mass function is truncated at low masses, the total 
mass of the cluster would not differ significantly from the estimate
of $10^{5}$ \msun\ in \citet{cla05} based on the un-modified Kroupa form. 
However, if the mass function is flat, or the optically-detected stars 
had a wider range of initial masses, the total mass of the cluster would 
be only 40,000-70,000 \msun. Obviously, an accurate measurement of the mass
function, and consequently the total mass, requires direct infrared 
observations of the low-mass stars in Westerlund 1. However, these
X-ray observations provide a useful starting point.

\subsection{Stellar Winds}

It is not clear whether stellar winds or supernovae are the dominant
source of the diffuse X-ray emission from Westerlund 1,
because it is at an age when both should contribute equally to its 
mechanical output 
(e.g., Leitherer, Robert, \& Drissen 1992)\nocite{lrd92}.
Individual stellar winds carry a power of 
\begin{equation}
L_{\rm w} = 3\times10^{35} 
\left( \frac{\dot{M}}{10^{-6}M_\odot/{\rm yr}} \right)
\left( \frac{v_{\rm w}}{10^3~{\rm km/s}} \right)^2~{\rm erg~s}^{-1},
\end{equation}
where $\dot{M}$ is the mass loss rate, and $v_{\rm w}$ is the 
wind velocity. 
The WR stars dominate the mechanical output from stellar winds, with typical 
$\dot{M} \approx 6\times10^{-5}$ \msun\ yr$^{-1}$ and 
$v_{\rm w} \approx 1700$ km s$^{-1}$, so that 
$L_{\rm w} \approx 5\times10^{37}$ \ergs\ \citep{lrd92}. With at 
24 WR stars in the cluster, the total mechanical energy output from winds
is $>1\times10^{39}$ \ergs. 

We can estimate the X-ray luminosity of the resulting cluster wind using the 
analytic solutions to the density and temperature that 
\citet[][see also Stevens \& Hartwell 2003]{crr00} 
derived for a wind expanding
supersonically into the interstellar medium \citep{cc85}. Within the
radius of the cluster where the stars input their energy, the density is
given by 
\begin{equation}
n_0 = 0.1 N 
 \left( \frac{\dot{M}_{\rm w}}{10^{-5} M_\odot/{\rm yr}} \right)
 \left(\frac{v_{\rm w}}{10^3~{\rm km/s}} \right)^{-1} 
 \left( \frac{R_c}{{\rm pc}} \right)^{-2}{\rm cm^{-3}}
\end{equation}
(note that we used the supersonic solution with an adiabatic index 
$\gamma = 5/3$ for the original equation) and the temperature by
\begin{equation}
kT_0 = 1.3\left( {{v_{\rm w}}\over{1000~{\rm km/s}}} \right)^2~{\rm keV},
\end{equation}
where $N$ is the number of stars, $\dot{M}_{\rm w}$ is the average mass 
loss rate of the stars, and $R_c$ is the radius within which the stars 
are contained and the 
winds are thermalized. If we use the values above for WR stars, 
assume $N$=24, and take the radius of the cluster to be $R_c$=4 pc 
($\approx$3\arcmin at 5 kpc), then we find $n_0$=0.6 cm$^{-3}$ and 
$kT_0$=4 keV. Using any of the standard plasma models in XSPEC 
\citep[e.g.,][]{mlo86,log95}, and converting $n_0$ and $R_c$ to an emission 
measure (i.e., $K_{\rm EM} = \frac{4}{3} \pi R_c^3 n_0^2$) we find a predicted 
$L_{\rm X} = 3\times10^{34}$ \ergs\ (2--8 keV). Therefore, a cluster wind
could in principle account for all of the diffuse X-rays from Westerlund 1.

However, the cluster wind model is not able to account for the spatial 
distribution of the diffuse X-rays from Westerlund 1. A cluster wind
would produce almost all of the diffuse X-ray emission within the core 
radius $R_c$ \citep{sh03}, whereas at 70\% of the diffuse emission from 
Westerlund 1 is part of a halo that extends out to at least 5\arcmin\ 
(Fig. \ref{fig:diffdist}). If consider only the core of the diffuse emission
as originating from stellar winds, then the cluster is underluminous by a 
factor of two.

This result is 
particularly surprising given that the standard cluster wind model applied
by \citet{sh03} to NGC 3603, R136, and NGC 346 predicted significantly
{\it less} flux than is observed. In those cases, \citet{sh03} favored
the hypothesis that cold material was being entrained in the wind. To 
reconcile our results for Westerlund 1 to the cluster wind model, we would 
have to assume that enough cold material is entrained that the cluster wind 
no longer emits in the 2--8 keV bandpass. This requires that the plasma
be cooled by a factor of $\approx$10, to $\la$0.4 keV. Based on Figure~1
in \citet{sh03}, we estimate that this would require that the mass of 
cold material input into the wind is twice that of the hot material,
or roughly $3\times10^{-3}$ \msun\ yr$^{-1}$. The mass loss rates from
the $\approx$10 luminous blue variables, red supergiants, and yellow
hypergiants, which should be $<$$10^{-4}$ \msun\ yr$^{-1}$ each
\citep[e.g.,][]{jk90,lrd92,svk04},
probably could not account for this large amount of cool mass. 
Therefore, either there is a currently-unseen source of mass in Westerlund
1, or the stellar winds are not thermalized within the cluster and escape 
without radiating much.


\subsection{Supernovae}

The presence of an isolated X-ray pulsar in Westerlund 1 \citep{mun06} 
confirms that supernovae have occurred there. If we extrapolate 
an initial mass function with slope $\alpha$=1.8--2.7
for $M$$>$30\msun\ to higher masses, we expect that the cluster originally 
contained $\approx$80--150 stars with initial masses $>$50\msun\ that have 
already undergone supernova. 
For the most massive stars, this would have started when the 
cluster was about 3~Myr old, so the
average supernova rate over the last 1 Myr should be on order one
every 7,000--13,000 yr. If each supernova had a kinetic energy of 
$10^{51}$ erg, then the average power released by supernovae is 
$\sim (2-5)\times10^{39}$ \ergs. 

No obvious supernova remnant is present in our \chandra\ image
of Westerlund 1, but this is not surprising. 
We have examined images from the {\it Spitzer} GLIMPSE program 
(R. Indebetouw, private communication), and 
there is no evidence that dense gas or dust still surrounds Westerlund 1.
Therefore, Westerlund 1 appears to have cleared away the ISM for parsecs 
around. When a supernova
occurs in such an evacuated cavity, a typical radio and X-ray remnant
is not expected until the remnant encounters the boundaries of the 
bubble blown by the cluster \citep[e.g.,][]{cde89}.

Whether the hard, possibly non-thermal emission from Westerlund 1 is 
produced by supernovae is unclear.
Most supernova remnants produce thermal X-ray emission with strong lines, 
but a few are also non-thermal X-ray sources. For example, RCW 86 and SN 1006 
exhibit non-thermal filaments near the outer boundary of the shock, 
and thermal emission in the interior \citep[e.g.,][]{drb04,rho02}. 
AX J1843.8$-$0352 and G346.3-0.5 exhibit non-thermal emission almost 
exclusively throughout the remnant \citep[e.g.,][]{uen03,laz05,hir05}. 
Unfortunately, there is not a satisfactory explanation as to why 
a small fraction of supernova remnants produce  
non-thermal emission, so the issue remains unresolved for Westerlund 1.

\subsection{Non-Thermal Particles\label{sec:nonthermal}}

In principle, non-thermal particles can be produced either in supernova 
remnants 
\citep[e.g.,][]{lp04} or in colliding stellar winds \citep[e.g.,][]{eu93}.
Once they are produced, inverse-Compton scattering should dominate 
synchrotron losses by a large factor in Westerlund 1  
\citep[see, e.g.,][]{rl79}. The ratio of the energy-loss rates is 
given by the ratio of the background radiation to the magnetic 
energy density. The energy density of the stellar light from 
the OB and WR stars in Westerlund 1 is approximately
\begin{eqnarray}
\nonumber U_{\rm phot} & = & {{L_{\rm stars}}\over{4\pi c D^2}} \\ 
 & = & 5.5\times10^{-9} \left({{L_{\rm stars}} \over {10^{7} L_\odot}}\right) \left(\frac{d}{1~{\rm pc}}\right)^{-2}~{\rm erg~cm}^{-3}
\end{eqnarray}
where $L_{\rm stars} \sim 10^7 L_\odot$ the luminosity of the 
cluster. 
For synchrotron losses to be important, magnetic fields would 
have to have an energy of $B^2/8\pi \ga U_{\rm phot}$, which corresponds
to $B\ga$0.4 mGauss. This is much stronger than the microGauss fields 
generally assumed for the interstellar medium \citep[e.g.,][]{beck01},
so inverse-Compton scattering is probably the dominant loss mechanism
for non-thermal particles.

If the non-thermal X-ray emission is produced by inverse-Compton
scattering, the
energy requirements are modest. Non-thermal particles would only
need to be replenished at a rate sufficient to balance the X-ray
luminosity, $3\times10^{34}$ \ergs. Furthermore, inverse-Compton 
scattering photons from 
optical and UV energies ($E_{\rm in} = 2-20$ eV) into the X-ray band 
($E_{\rm out} \approx 3$ keV) only requires electrons with 
$\gamma^2 \sim E_{\rm out}/E_{\rm in}$, or energies of 6--20 MeV. 
These particle energies are rather small. 
For comparison, if the magnetic field in Westerlund 1 has a strength of 
only 10 $\mu$Gauss, producing non-thermal synchrotron radio emission like 
that seen from 30 Dor C \citep{bam04} requires electrons with energies of 
a few GeV. Therefore,
detecting diffuse, non-thermal radio emission from a star cluster like
Westerlund 1 \citep[e.g.,][]{yz03} would provide a much more interesting 
constraint on the maximum 
energies of particles than detecting non-thermal X-rays. The interferometric
radio observations in the literature \citep{cla98} would have resolved
out arcminute-scale diffuse radio emission, so single-dish observations 
are necessary to determine whether higher-energy particles are also produced
by the cluster.






\section{Summary}

We have identified diffuse X-ray emission within 5\arcmin\ of the core
of Westerlund 1 with a modest luminosity of $(3\pm1)\times10^{34}$ \ergs\ 
(2--8 keV). This low luminosity is puzzling, because unresolved 
pre-main-sequence stars, 
a thermalized cluster wind, or a series of supernova remnants would each
be expected to produce at least this much X-ray emission.
Therefore, one or all of these mechanisms is not producing nearly as
much X-ray flux as would be expected based on comparison with other 
star clusters and with theoretical calculations.

The lack of a 6.7 keV He-like Fe line accompanying the hard 4--8 keV 
continuum implies that no more than 30\% of the diffuse emission
is produced by young stellar objects. Therefore, we infer that there
are $\la$40,000 stars with masses between 0.3 and 2 \msun, which is 
significantly fewer than the $10^{5}$ stars one would expect from 
extrapolating the number of massive, optically-identified stars 
to lower masses using a standard initial mass function \citep{cla05}. 
Moreover, this 
limit is conservative, because in computing it we have assumed that the
line emission from the low-mass stars would be produced by a solar 
abundance of iron. If we had assumed that iron had an abundance several times
the solar value, as it does in the spectra of stars in Orion \citep{fei05},
then only a few percent of the diffuse 2--8 keV flux could be produced by 
pre-main sequence stars.

In contrast, the lack of iron emission in the spectrum is consistent
with a similar under-abundance of iron that is observed in X-ray spectra 
of individual O and WR stars \citep[e.g.,][]{ski01,ski02a,sch03,ski05}. 
However, if the O and WR star winds collide and
thermalize as expected, they would form a pressure-driven cluster 
wind that would expand and cool rapidly \citep{crr00,sh03}. Such a 
wind would not radiate in 
the X-ray band outside of the cluster core, and therefore cannot
explain the broad halo of emission between $\approx$3\arcmin\ and 5\arcmin. 

Instead, the halo 
of X-rays may represent non-thermal particles accelerated by the 
colliding stellar winds or by supernova remnants. 
However, the energy lost in X-rays represents less than 
$10^{-5}$ of the kinetic energy released by stellar winds and 
supernova remnants. The rest of the energy either
(1) emerges below 2 keV where our observations are insensitive, 
(2) dissipates beyond the bounds of 
our image ($\approx$7 from the cluster core) when the cluster wind or 
supernova remnants impact the ISM, or (3)
escapes the Galactic plane, enriching the intergalactic medium with metals. 
We plan to address the second option in the next year, by observing a larger 
area around the cluster ($\approx$15\arcmin) with \xmm. 

\acknowledgments
We thank E. Feigelson and an anonymous referee for comments and 
suggestions that improved this manuscript.
Support for this work was provided by the National Aeronautics and Space 
Administration through Chandra Award Number 06400311 issued by the Chandra 
X-ray Observatory Center, which is operated by the Smithsonian Astrophysical 
Observatory for and on behalf of the National Aeronautics Space 
Administration under contract NAS8-03060.
Support for MPM was provided by NASA through Hubble Fellowship grant 
HST-HF-01164.01-A awarded by the Space Telescope Science Institute, which is 
operated by the Association of Universities for Research in 
Astronomy, Inc., for NASA, under contract NAS 5-26555

\begin{small}

\end{small}

\end{document}